\documentclass[onecolumn]{aa} 
\usepackage{amsmath} 
\usepackage[pdftex]{graphicx} 
\usepackage{txfonts}
\usepackage{url}
\usepackage{longtable}
\begin{document}
\title{Short-term variability of comet C/2012 S1 (ISON) at 4.8 AU from the Sun}


   \author{P. Santos-Sanz\inst{1}
		   \and	
		   J.L. Ortiz\inst{1}
		   \and
		   N. Morales\inst{1}
		   \and
		   R. Duffard\inst{1}
		   \and
		   F. Pozuelos\inst{1}
		   \and
		   F. Moreno\inst{1}
		   \and
		   E. Fern\'{a}ndez-Valenzuela\inst{1}
}

\offprints{P. Santos-Sanz: psantos@iaa.es}

\institute{Instituto de Astrof\'{i}sica de Andaluc\'{i}a-CSIC, Glorieta de la Astronom\'{i}a s/n, 
18008-Granada, Spain.
 \email{psantos@iaa.es} 
}
   
   \date{Received 3 November 2014 / Accepted 26 December 2014}

 
  \abstract 
   {We observed comet C/2012 S1 (ISON) during six nights in February 2013 when it was at 4.8 AU from the sun. At this distance and time the comet was not very active and it was theoretically possible to detect photometric variations likely due to the rotation of the cometary nucleus.}
   {The goal of this work is to obtain differential photometry of the comet inner coma using different aperture radii in order to derive a possible rotational period.}
   {Large field of view images were obtained with a 4k $\times$ 4k CCD at the f/3 0.77m telescope of La Hita Observatory in Spain. Aperture photometry was performed in order to get relative magnitude variation versus time. Using calibrated star fields we also obtained ISON's R-magnitudes versus time. We applied a Lomb-Scargle periodogram analysis to get possible periodicities for the observed brightness variations, directly related with the rotation of the cometary nucleus.}
   {The comet light curve obtained is very shallow, with a peak-to-peak amplitude of 0.03 $\pm$ 0.02 mag. A tentative synodic rotational period (single--peaked) of 14.4 $\pm$ 1.2 hours for ISON's nucleus is obtained from our analysis, but there are other possibilities. We studied the possible effect of the seeing variations in the obtained periodicities during the same night, and from night to night. These seeing variations had no effect on the derived periodicity. We discuss and interpret all possible solutions for the rotational period of ISON's nucleus. }
   {}

   \keywords{Comets: individual: C/2012 S1 (ISON) -- Optical: planetary systems -- Methods: observational  -- Techniques: photometric}
   
\titlerunning{Short-term variability of comet C/2012 S1 (ISON) at r = 4.8 AU}

   \maketitle

%

\section{Introduction}
\label{intro}

 Comet C/2012 S1 (hereafter ISON) was discovered at 6.3 AU from the Sun on September 21, 2012 by V. Nevskiand and A. Novichonok using the International Scientific Optical Network (ISON) near Kislovodsk, Russia (\cite{ison_discover}). The orbit of the comet was quickly computed using precovery images from the Mount Lemmon Survey and Pan-STARRS facilities. The computed orbit was nearly parabolic, which indicated that it was a dynamically new comet coming directly from the Oort cloud, with a nucleus probably including plenty of fresh ices and species (\cite{2014A&A...564L...2A}) never irradiated by the Sun, like the nucleus of comet C/1995 O1 (Hale-Bopp; \cite{1997EM&P...77.....A}). Because of this, a huge outgassing near the perihelion was expected, with a large rate of ice sublimation and dust dragged by the activity from the cometary nucleus. Unfortunately, models that predict the behaviour of a new Oort cloud comet produce large uncertainties when the comet is approaching its perihelion because usually there is no a priori knowledge of many relevant parameters needed for the modelling, and finally ISON's activity close to its perihelion was not so huge as was expected by some initial predictions.
 
ISON was a sungrazing comet which reached its perihelion on November 28, 2013, when it passed at only 0.012 AU (2.7 R$\odot$) from the Sun with a V magnitude of $\sim$ -2 mag according to SOHO/STEREO images (\cite{ison_end}). This very close approach suggested the possibility that the comet could disintegrate near its perihelion, which actually happened: ISON's nucleus did not survive the close approach to the Sun and it totally vanished (\cite{2014arXiv1404.5968S}). At the first moment after the perihelion passage, the nucleus --or at least chunks of particles from the nucleus-- seemed to have survived, but finally the brightness of ISON faded dramatically in a short period of time (\cite{ison_end}; \cite{ison_end2}; \cite{moreno2014}). Numerous amateur and professional astronomers tried to recover the comet some days after the perihelion; however, the results were negative (\cite{ison_no_recover}).

Rotational periods of several cometary nucleii have been obtained by means of time series photometry (\cite{1978Icar...34....1F}; \cite{1990ApJ...351..277J}; \cite{1997A&A...326.1268M}; \cite{2003A&A...407L..37G}; \cite{2005A&A...444..287S}; \cite{lamy2005}; \cite{2008MNRAS.385..737S}; \cite{2012A&A...548A..12L}; \cite{2014A&A...569L...2M}). This task is easier when the comet nucleus is not active and there is not a gas/dust coma (i.e. only the bare nucleus), but it is even possible when there is activity and the comet presents a coma. When a comet nucleus is active, the brightness measured in a region close to it will show three types of photometric variability: i) variability due to the rotation of a non-spherical nucleus which reflects different amounts of sunlight depending on the cross section; ii) periodic variations of the gas/dust production due to diurnal insolation; and iii) activity outbursts and/or other random changes. So, ideally, to obtain the variability due to the rotation of the nucleus it is best to observe when there is no coma, but in case of activity it is also possible to obtain the rotational period of the nucleus by means of the photometry of the inner coma which can inform us about cyclic diurnal activity variations linked to the rotation of the nucleus (\cite{1986Natur.324..646M}; \cite{1990AJ....100..896S}). 

It is also known that the measured amplitude of the variations may depend on the aperture size used (\cite{1993AJ....106.1222M}). Another factor to take into account are the seeing variations during each night and from night to night which might affect the final photometry and might introduce spurious periodicities (\cite{1990ApJ...351..277J}; \cite{2000AJ....119.3133L}). In our ISON observations at $r_{h}\sim 4.8$ AU the comet had a coma, but it was in a quiescent state. This allowed us to derive the rotational period by means of aperture photometry. We also analysed the effect in the photometry of the mentioned seeing variations during our observations.

In Sect. \ref{observations} we present the information about the acquisition of ISON's images during the observing run. The data reduction and the photometric technique applied are described in Sect. \ref{datareduction}. The results obtained for the rotational period of ISON's nucleus, after the application of the Lomb-Scargle periodogram technique, are detailed in Sect. \ref{results} and discussed in Sect. \ref{discussion}. Finally, a summary of the main results of this work is presented in Sect. \ref{summary}.


\section{Observations}
\label{observations}

ISON was observed for almost six nights in a row on February 8-9 and 11-14, 2013. Images of the comet were obtained from the f/3 0.77m telescope located at La Hita Observatory in Toledo, Spain, for several hours each night, when it was at heliocentric distances (r$_{h}$ ) of 4.84-4.77 AU and at geocentric distance ($\Delta$) of 4.01 AU. The detector used was a 4k $\times$ 4k CCD camera which provided a field of view (FOV) of $48.1' \times 48.1'$ with a 0.705 arcsec/pixel scale of image. This scale is enough to provide accurate point spread function (PSF) sampling, even in the best seeing cases. The median seeing during the observations was 4.8$''$ (see Table \ref{tableOBS}).

Integration time was chosen taking into account two factors: i) it had to be sufficiently long to achieve a high enough signal-to-noise ratio (S/N) to study the comet; and ii) it had to be short enough to avoid elongated images of the comet (because the telescope tracked at sidereal speed in order to use the stars in the field to do relative photometry). Considering these two criteria, and the poor median seeing, the integration time chosen was 150 seconds. No binning was used for the image acquisition. During the observing dates the comet was moving at a median speed $\sim$ 0.55 arcsec/min, so it spanned around 2 pixels in the images during the total integration time (i.e. the comet trail is negligible compared with the median seeing, and differential tracking is not needed). Images of the comet were acquired without filter in order to maximize the S/N.

Since the goal of this work is to study the brightness variability of comet ISON, we use relative photometry, not absolute photometry, to search for possible short-term periodicities. Our instrument setup is ideal in this regard because the very large FOV allows us to always use the same reference stars for photometry, at least for several days, which permits a higher photometric precision than absolute photometry. The dates of observations, number of ISON images, geometric data, and night conditions during the observing run are summarized in Table \ref{tableOBS}.

\begin{table*}[!hbt]

\caption{Observational circumstances of comet C/2012 S1 (ISON) during February 2013} 

\label{tableOBS} 

\centering 

\begin{tabular}{l c c c c c c} 

\hline\hline 

Date	    & JD 	    &		images &	r$_{h}$ &	$\Delta	$	&	$\alpha$ &	FWHM	\\
		&	[days]	&		       &	[AU]	&	[AU]	    &   [deg]  & [arcsec]	\\

\hline 

2013-Feb-08	    & 2456332.47479-2456332.58176 &	42    &	4.839   &  4.010  	&   6.963	 &  4.87	\\		
2013-Feb-09	    & 2456333.37477-2456333.62491 &	97    &	4.828   &  4.009  	&   7.169    &  4.65	\\		
2013-Feb-11	    & 2456335.38203-2456335.61912 & 68    &	4.806   &  4.008  	&   7.587	 &  5.30	\\		
2013-Feb-12	    & 2456336.40498-2456336.50574 & 40    &	4.796   &  4.008  	&   7.780	 &  5.01	\\		
2013-Feb-13	    & 2456337.42286-2456337.60275 &	76    &	4.784   &  4.008  	&   7.996	 &  4.59	\\		
2013-Feb-14	    & 2456338.32476-2456338.60417 &	79    &	4.773   &  4.008  	&   8.188	 &  4.14	\\

\hline 

\end{tabular}

\begin{flushleft}
\footnotesize{\textbf{Date} is the date of observation; \textbf{JD} is the first and last Julian dates for each observing night; \textbf{images} is the number of useful images; \textbf{$r_h$ [AU]:} heliocentric distance (AU); \textbf{$\Delta$ [AU]:} geocentric distance (AU); \textbf{$\alpha$[deg]:} phase angle (degrees); \textbf{FWHM:} median of the full width at half maximum of the seeing for each observing night (arcseconds).}
\end{flushleft}

\end{table*}


\section{Data reduction}
\label{datareduction}

Series of sky flatfield and bias images were obtained during each observing night to correct the images of the comet. Using these images we built a median bias and a median flatfield for each observing night. We take enough bias and flatfield images so that we can reject those that might be affected by acquisition or observational problems. Each day a median flatfield was constructed using twilight dithered images. The final median flatfield image was inspected for possible residuals from very bright or saturated sources. The flatfield exposure times were long enough to avoid any shutter effect. Each comet image was bias subtracted and flatfielded using the median bias and median flatfield in order to remove image gradients and artefacts. No cosmic ray removal algorithms were used but images with cosmic rays or stars very close to the position of the comet were rejected in order to get the highest quality and uncontaminated photometry.


\subsection{Photometry}
\label{photometry}

Daophot routines (\cite{1987PASP...99..191S}) were used to perform relative photometry using a maximum of 25 stars. The mean error bar of the individual relative magnitudes of the comet was $\sim$ 0.02 mag. Special care was taken not to introduce bad results due to background sources in the aperture or in the background annulus chosen to compute the sky level. Images affected by cosmic ray hits or artefacts within the flux aperture are not included in our final photometry.

\begin{figure}[!hpbt]
   \centering
   
   \includegraphics[width=9cm]{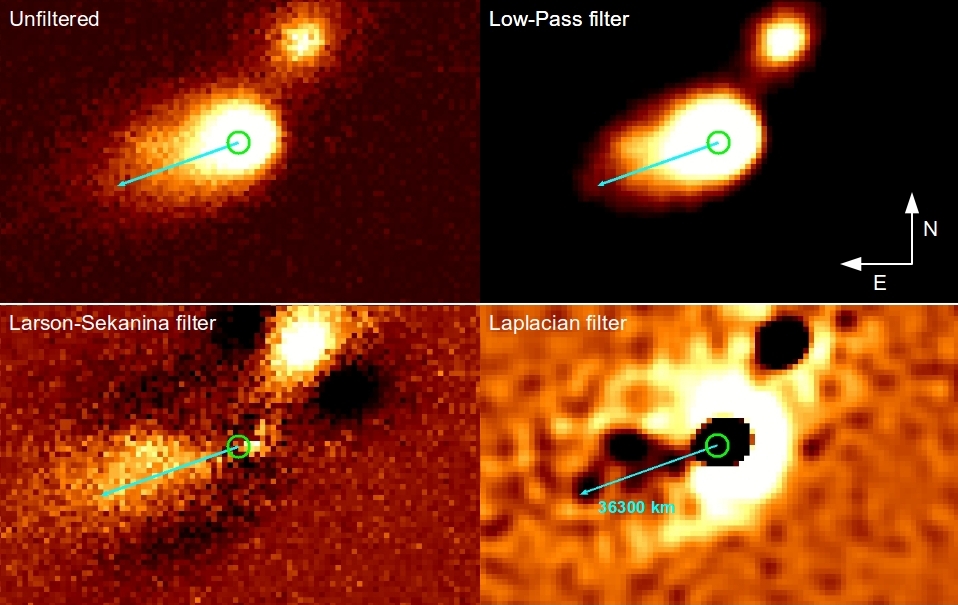}
   
   \includegraphics[width=9cm]{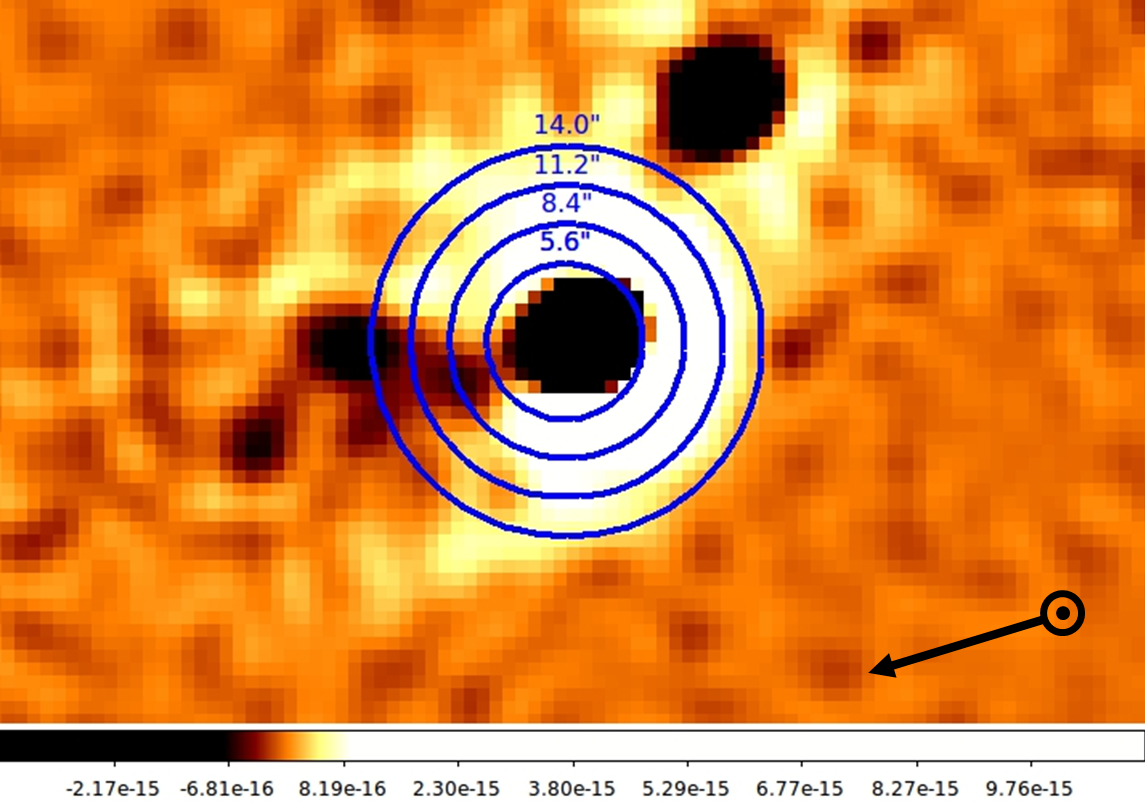}
  
      \caption{ISON image from the 1.52 m telescope at Sierra Nevada Observatory taken on February 2013 14.04 UT. \textbf{In the upper four panels} the unfiltered image (upper left) and filtered images using different digital techniques: Low Pass, Larson-Sekanina, and Laplacian filters (\cite{2014Icar..239..168S}). Green circles mark the comet photocentre. Green arrows indicate the small tail or coma enhancement structure pointing to the antisolar direction, clearer in the filtered images, in particular in the Larson-Sekanina and Laplacian ones. \textbf{In the bottom panel}, zoom-in of the ISON image processed with the Laplacian filter. Blue circles are the different apertures (diameter in arcseconds) used to perform the photometry. The smallest (aperture = 5.6$''$) is not contaminated by the small antisunward tail or coma structure. In all panels north is up and east to the left. The antisolar direction is shown with a black arrow.
}
      
         \label{Ison_filtered}

\end{figure}

A common reduction software programmed in IDL was used for the photometry of all the previously calibrated images. Trial aperture sizes ranged from 8 to 20 pixels. In general, it was necessary to
choose an aperture that was as small as possible to get the highest S/N, but large enough to measure
most of the flux. Then we needed to use an aperture radius that was about the same size as the FWHM of the
seeing, and to check several aperture radii values around this one. In the case
of a comet, even if it is at moderately large heliocentric distances (as was comet ISON when we observed
it), activity is expected (i.e. at 4.8 AU no high activity is expected, at least no H2O sublimation,
only activity triggered by CO or CO2 sublimation). With all this in mind, we finally found the aperture that gave the lowest dispersion in the photometry of the comet and of the stars with similar brightness. In our particular case we also used a deeper and better space-resolved image of the comet inner coma, acquired with a larger telescope, to help us select the optimum aperture radius. This image was acquired in February 2013, 14.04 UT (\cite{moreno2014}), very close to the dates of our observations. It was taken with the 1.52 m telescope located at Sierra Nevada Observatory and operated by the Instituto de Astrof\'{i}sica de Andaluc\'{i}a (CSIC) in Spain. We used a 2048$\times$2048 VersArray CCD attached to the telescope in a 2$\times$2 binning mode providing a scale of image of 0.46 arcsec/pixel and a red Johnson-Cousins filter. The seeing during this observing night was 1.63$''$. The image was processed and filtered with different digital techniques (\cite{2014Icar..239..168S}) in order to enhance structures in the inner coma; the results of this processing are shown in the four upper panels of Figure \ref{Ison_filtered}. A small tail structure or coma enhancement pointing to the antisolar direction is clearly visible in the Larson-Sekanina and Laplacian filtered images. We placed different apertures at the photocentre of the Laplacian filtered image in order to see which aperture radii are affected by the tail or coma structure. As we can see in the figure (blue circles in the bottom panel of Figure \ref{Ison_filtered}), the antisunward tail or coma structure is avoided for aperture diameters less than or equal to $5.6''$ (i.e. 8 pixels for the plate scale of La Hita Observatory images). We selected this aperture diameter as the best one in order to perform the photometry: it is large enough to include a significant amount of flux (FWHM$_{median} = 4.8''$) and small enough to avoid the antisunward tail or coma structure.

Several sets of reference stars were used to get the relative photometry, but only the set that gave the lowest scatter was chosen. In some cases, several stars per field had to be rejected from the analysis because they exhibited some variability. The final photometry of ISON was computed taking the median of all the light curves obtained with respect to each reference star. This median technique eliminates spurious results and minimizes the scatter of the photometry.

To perform relative photometry of moving targets we tried, ideally, to keep the same reference stars during the whole campaign. As ISON was not a slow moving target during the observation dates (speed$\sim$ 0.55 arcsec/min), it was not possible to keep the same reference stars during the whole campaign. Nevertheless, because the FOV of the instrument is nearly one degree we were able to use the same reference stars on at least two or three nights, which allows us to obtain high accuracy relative photometry. We link the time series photometry from different sets of nights using images of the same fields acquired under photometric conditions after the observing campaign. We took four images for each night field for the six different date fields during the same photometric night. We used these images to calibrate in flux using the same set of reference stars used to perform the relative photometry of the comet. We also checked the stability of this method using different aperture sizes. Finally, we used the calibration factors obtained from this technique to link the time series photometry of nights with different reference stars in order to avoid small jumps in the photometry.

In Sect. \ref{observations} we mentioned that we acquired the comet images with no filter. Using no filter may produce strong fringing effects caused by near-infrared interferences. Luckily, this was not the case for our CCD detector. On the other hand, as we mentioned, obtaining unfiltered images allowed us to reach deeper magnitudes with enough S/N. 

\section{Results}
\label{results}

The time series photometry of the comet obtained as explained in Sect. \ref{photometry} was inspected for periodicities using the Lomb technique (\cite{1976Ap&SS..39..447L}) programmed as described in \cite{1992nrfa.book.....P}. The photometry needs some corrections before applying the periodogram analysis: i) first we corrected the Julian dates by subtracting the one-way light-time, which is the elapsed time since light left the comet (i.e. we referenced the time to comet-centre coordinates instead of topocentric ones). Light-time correction during the February 2013 ISON campaign was typically $\sim$ 33 minutes; ii) we also corrected the relative magnitudes for the changing heliocentric (r$_{h}$) and geocentric distances ($\Delta$) by means of the expression 

\begin{equation} m_{R}(1,1) = -2.5 \cdot log(flux) - n \cdot log(r_h) - 5 \cdot log(\Delta),
\label{equation}
\end{equation}
where $m_{R}(1,1)$ is the corrected (by r$_{h}$ and $\Delta$, expressed in astronomical units --AU) relative magnitude, $flux$ is the relative measured flux in ADUs (analogue-to-digital units), and $n$ is an index which reflects the change of the magnitude with the heliocentric distance. This index is usually 5 for asteroids and airless bodies, and can be larger for objects with an atmosphere or an active coma, like active comets. We tried different values for $n$, from $n = 5$ (i.e. asteroidal behaviour) to $n = 20$ (very active coma or atmosphere). For all these values we obtained compatible results from the Lomb periodogram analysis. Finally we chose $n = 5$ as the preferred value because it implies an asteroidal behaviour, but it is also compatible with a quiescent or stable coma with a nearly constant amount of dust inside the photometric aperture. This choice is supported by the magnitude slope reported by \cite{2014arXiv1404.5968S} for the time of our observations (see Figure 2 of the cited work) when the magnitude changed as r$_{h}^{-5}$, which means $n$ = 5. 

\begin{figure}[!hpbt]
   \centering
   
   \includegraphics[width=9.4cm]{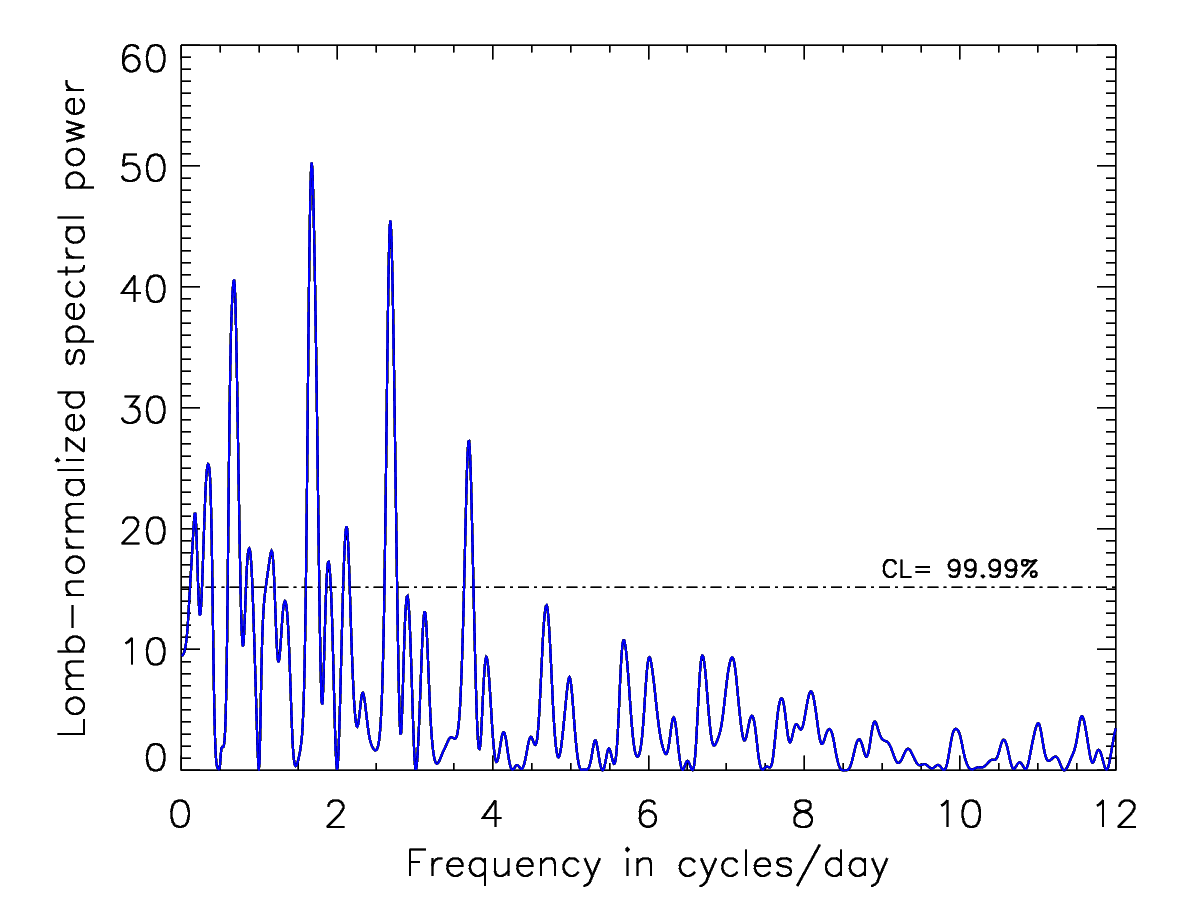}

      \caption{Lomb-Scargle periodogram from ISON data obtained by means of aperture photometry with an aperture size of 5.6$''$, and corrected by heliocentric and geocentric distances using Eq. \ref{equation}. The peak (global maximum) at 1.67 cycles/day with normalized spectral power equal to 50.25 corresponds to a synodic-rotation-period of 14.4 $\pm$ 1.2 hours, which we take as our preferred rotational period (see Sect. \ref{period} for a detailed explanation). The confidence level above the horizontal dashed line is 99.99\%.
}
      
         \label{lomb}

\end{figure}

The periodogram obtained after the analysis of these data is shown in Figure \ref{lomb} and is analysed in Sect. \ref{period}. Similar periodograms were obtained using different photometric apertures and n-indexes, but we have justified our preferred aperture diameter (5.6$''$) in Sect. \ref{photometry} and our preferred n-index (n = 5) in the current section.

We have checked as well if the seeing variations from night to night and during the same night may have affected the derived photometry producing spurious periodicities as suggested in the work by Licandro et al. (2000). To discard this effect, we searched for possible correlations between the measured magnitudes and the FWHMs obtained for each image. We do not see any correlation of the relative magnitudes versus the FWHMs using a Spearman test (\cite{Spearman1904}). To be even more confident we search for time periodicities of the FWHMs using the Lomb technique (\cite{1976Ap&SS..39..447L}) and we do not get any reliable period, so we concluded that the brightness variabilities (and periodicities) observed in our photometric data are real (i.e. due to variability likely related with the rotation of the nucleus) and not due to variations on the weather conditions.

Apart from the relative photometry, we have also estimated absolute photometry using USNO-B1 stars in the FOV as photometric references. With this technique we obtained an estimation of the R magnitudes of comet ISON during our observing campaign, but the uncertainty in the absolute calibration is large, around 0.4 mag because the USNO-B1 magnitudes are not standard BVRI magnitudes and we are not using a filter. In Table \ref{photomresults} (online material) we show the whole time series photometry of comet ISON: dates and R magnitudes for each image. The mean R-magnitude obtained is 16.23 mag. This table also includes m$_R(1,1)$, which are the R magnitudes that ISON would have at 1 AU from the Sun and the Earth (no phase correction applied), and the geometrical circumstances of the comet during the observing dates, such as the phase angle ($\alpha$), and the heliocentric (r$_{h}$) and geocentric ($\Delta$) distances.

\subsection{Tentative rotational periods}
\label{period}

The main peak in the Lomb-Scargle periodogram (Figure \ref{lomb}) is centred at 1.67 cycles/day (14.4 h) with a normalized spectral power of 50.25. There are what appear to be two 24 h aliases at $\pm 1$ cycles/day from the main peak (at 9.0 h and 35.7 h respectively) with smaller spectral power than the primary one. We take 14.4 h as the real rotational period, although these other periods could also be possible, in particular 9.0 h, the one with the second largest spectral power. It is important to note that in time series photometry of low amplitude variability one can sometimes find that the true period corresponds to what appears to be an alias (\cite{2007A&A...468L..13O}; \cite{2010A&A...522A..93T}). We estimated the uncertainty in the periods by means of the FWHMs of the main peaks in the Lomb periodogram, and our preferred synodic rotational period is 14.4 $\pm$ 1.2 hours. This is a single-peaked rotational period, which means that the variability could be due to active region(s) distributed along the nucleus producing measurable (though very small) photometric variations in the inner coma. If variability were due to a rotating nucleus with non-spherical shape, the light curve should be double-peaked, and the rotational period should be double, i.e. 28.8 hours. In principle we take the single-peaked solution as the preferred one as is discussed in Sect. \ref{discussion}.

\begin{figure}[!hpbt]
   \centering
   
   \includegraphics[width=9.3cm]{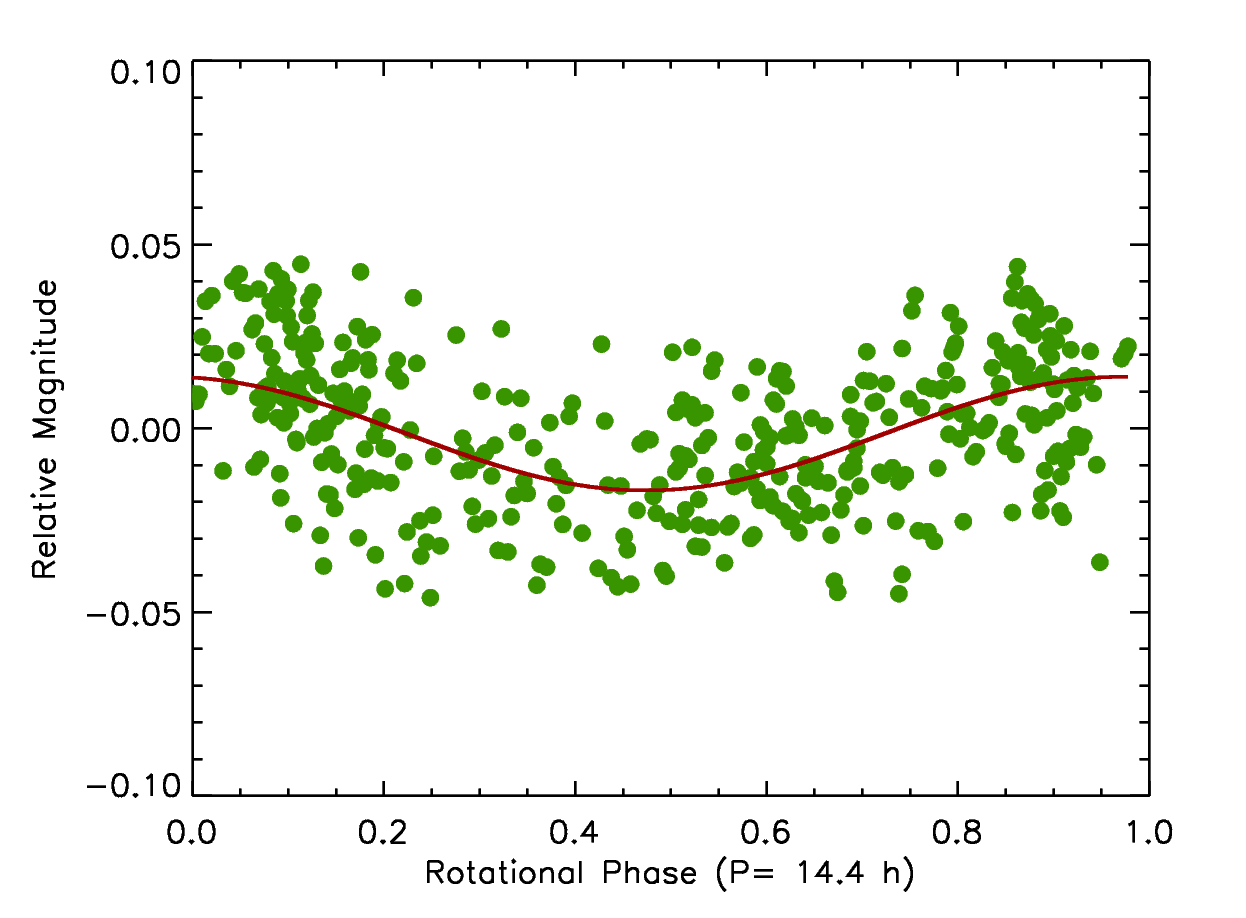}
   
   \includegraphics[width=9.3cm]{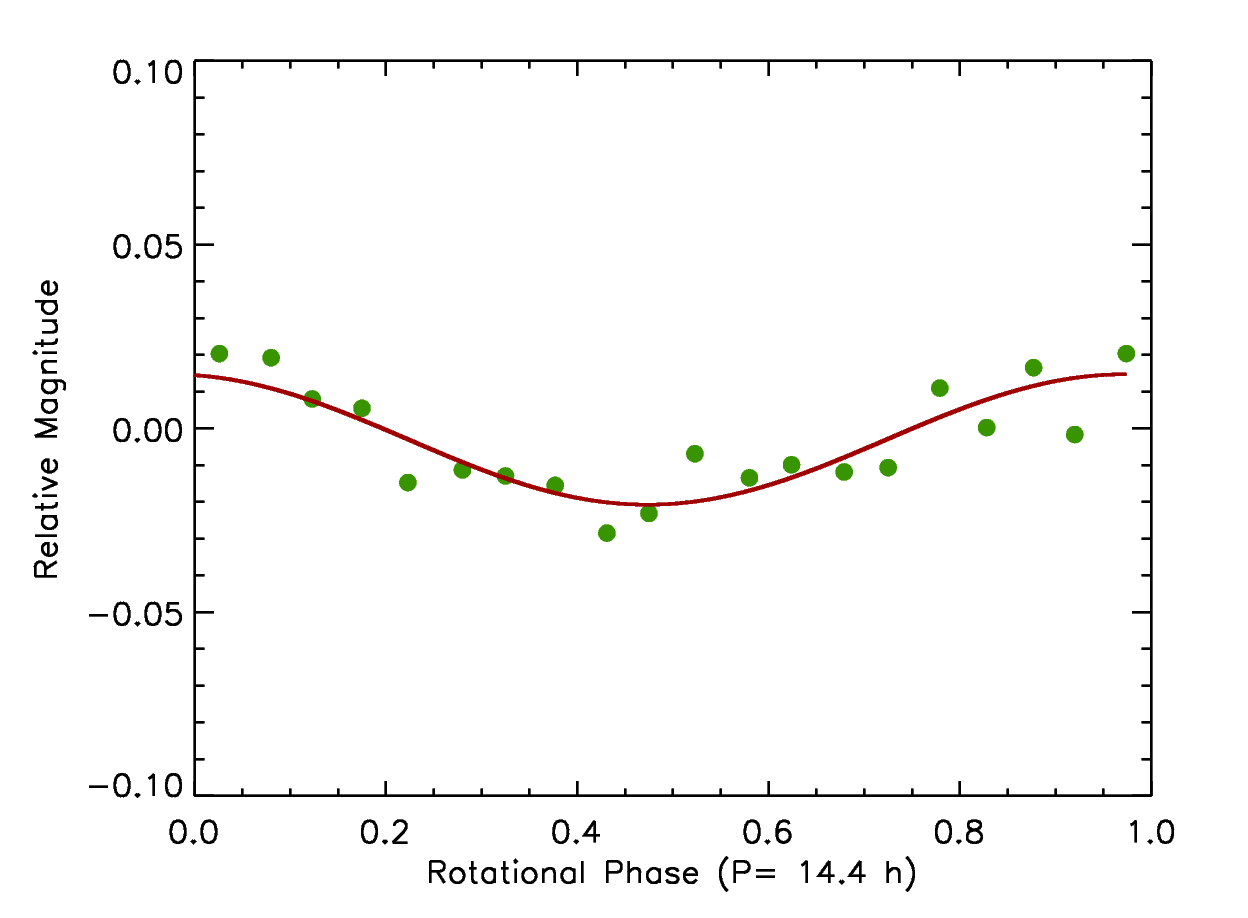}
   
     \caption{\textit{Single-peaked light curve of comet ISON.} \textbf{Upper panel:} Rotational light curve of comet ISON phasing all the photometric data to the preferred period (14.4 hours). Errors are not plotted for clarity, but each individual value has an uncertainty $\sim$ 0.021 mag. A one-term Fourier fit (red curve) is shown on top of the data, the peak-to-peak amplitude derived from this fit is 0.03 $\pm$ 0.02 mag. The Julian date for zero phase angle is 2456332.47479 days. \textbf{Lower panel:} Rotational light curve of comet ISON after applying a running median (0.05 step in phase) to the data showed in the upper panel. The rotational light curve variability is clearer to the eye in this plot. Like in the upper panel, a one-term Fourier fit (red curve) is shown on top of the data. Each individual point has an uncertainty $\sim$ 0.015 mag. }
      
         \label{Ison_LC}

\end{figure}

We phased the data described in the first paragraphs of Sect. \ref{results} using the preferred rotational period for ISON's nucleus (14.4 h) and taking 2456332.47479 days as the Julian date for zero phase angle (corresponding to the beginning of our observing run). The resulting rotational light curve is shown in the upper panel of Figure \ref{Ison_LC}. We fitted these data with a one-term Fourier function obtaining the red continuous line shown in the figure. From this Fourier fit we derived a peak-to-peak amplitude (i.e. the full amplitude) of 0.03 $\pm$ 0.02 mag. The bottom panel of Figure \ref{Ison_LC} is the result after applying a running median, with a 0.05 step in phase, to the upper panel light curve. ISON's light curve is cleaner and clearer in this running median curve (Figure \ref{Ison_LC}, bottom panel). The peak-to-peak amplitude derived from a Fourier fit to this median light curve is 0.04 $\pm$ 0.02 mag, slightly larger than that derived from the other fit, but compatible within the error bars.

We also built, plotted, and fitted in a similar manner the double-peaked light curve, phasing the data using the 28.8-hour period. The derived peak-to-peak amplitude in this case is exactly the same as for the single-peaked light curve: 0.03 $\pm$ 0.02 mag. We did not observe differences among the heights of the two peaks of the double-peaked light curve.

\section{Discussion}
\label{discussion}

The observations presented here were obtained from 293 to 287 days prior the perihelion, when comet ISON was not very active and the synthetic aperture CCD photometry technique can be used to extract information of the rotational period of the nucleus. In a recent work about this comet by \cite{2014arXiv1404.5968S} they identify five ignition points which define the comet's brightness/activity evolution from a date before discovery to perihelion. According to this work, our observations were acquired at the end of the second ignition point, when the comet was in a depletion stage, which they call Depletion Stage B, before the next ignition point (see Figure 1 in \cite{2014arXiv1404.5968S}). Taking into account these results, the comet nucleus was in a quiescent stage (i.e. no dramatic variation in brightness or in activity) when we acquired ISON's images used in this work. This result supports the conclusion that the variability in the coma photometry we reported is directly related with the rotation of the nucleus and not to activity or outgassing uncoupled from diurnal activity.

On the other hand, Moreno et al. (2014) and Li et al. (2013) claim that comet nucleus illumination geometry was almost constant until the last week before perihelion (i.e. the latitude of the subsolar point remains essentially constant until $\sim$ 1 AU). This means that almost the same face of the comet was illuminated by the sun until at least $r_{h}\le 1$ AU. They also proved, from dust models applied to February 14, 2013, images (\cite{moreno2014}), and by direct detection in Hubble Space Telescope (HST) images from April 10, 2013 (\cite{2013ApJ...779L...3L}), the existence of a jet pointing to the sunward direction with an opening angle $\sim 45^{\circ}$. The fact that the jet morphology did not change during HST and ground-based observations (\cite{2013DPS....4540701K}) and the small opening angle measured imply that it was a quasi polar jet which constrained the orientation of the nucleus spin axis. We did not detect this quasi-polar sunward jet due to the low resolution of our images. From these results they derived a large obliquity for the spin axis (I $\sim$ 70$^{\circ}$ in \cite{moreno2014}, and I = 50$^{\circ}$-80$^{\circ}$ in \cite{2013ApJ...779L...3L}). This large obliquity, jointly with the small phase angle at the dates of our observations ($\alpha_{mean} = 7.6^{\circ}$), points to a close to pole-on spin-axis orientation.

The very shallow light curve we calculated, with an amplitude well below 0.1 mag ($\Delta m$ = 0.03 $\pm$ 0.02 mag), could be due to i) a non-spherical nucleus rotating with a small aspect angle, close to a pole-on geometry (i.e. with a spin axis pointing approximately towards Earth) whose changing cross section would induce the observed variability, and/or ii) could be produced by a rotating nucleus with different active region(s) distributed along the surface which modulate the inner coma brightness. A combination of i) and ii) could also produce the rotational light-curve that we detected.

We did some calculations in order to quantify the percentage of the amplitude due to a non-spherical rotating nucleus embedded into the coma. Assuming an effective size of 4 km for the nucleus (\cite{2013ApJ...779L...3L} estimates an upper limit of 4 km for the effective nuclear diameter of ISON) we obtained that a bare nucleus with this effective diameter and a geometric albedo of 5\% at $r_h =$ 4.8 AU and $\Delta$ = 4.0 AU would have an apparent R-magnitude $\sim$ 21.9 mag. This means that the nucleus would be $\sim$ 190 times fainter than the measured mean R-magnitude of the coma (R $\sim$ 16.2 mag, from the online Table \ref{photomresults}); in other words, the contribution of the bare nucleus to the total flux (coma+nucleus) would be $\sim$ 0.5\%. With this in mind we estimated the percentage of variation contributed by the bare nucleus (R $\sim$ 21.9 mag) to the total flux (R $\sim$ 16.2 mag), assuming a highly elongated nucleus with an axes ratio of 2 (i.e. assuming a Jacobi ellipsoid shape for the nucleus with axes: $a>b>c$, where $c$ is the spin axis, and $a/b = 2$), and an aspect angle of 90$^{\circ}$ (which we know was not real because it was close to 0$^{\circ}$). Under these assumptions the peak-to-peak amplitude due to the sunlight reflected by such rotating nucleus embedded into the coma is $\sim$ 0.003 mag, which is around 10\% of the measured variability of ISON ($\Delta$m = 0.03 $\pm$ 0.02 mag). As the effective diameter of ISON's nucleus was likely $<$ 4 km (as stated by other authors like \cite{2013CBET.3720....1D} who estimated D $<$ 1.2 km), the axes ratio (a/b) was probably $<$ 2, and the aspect angle was much smaller than $90^{\circ}$, these estimations are upper limits to the variability produced by the shape of the nucleus. So we conclude that the contribution of the reflected sunlight in the rotating nucleus to the total flux variation would be even smaller than 10\%. We can conclude that the variability we measured in the inner coma is probably mainly due to active region(s) in the nucleus with a tiny and undetectable contribution due to the reflected sunlight in the rotating bare nucleus. These active region(s) located in the rotating nucleus would produce the measured variability, direcly related with the rotational period.

Taking into account the results from the literature we can draw the following scenario during our February 2013 photometric campaign: a cometary nucleus with one of its hemispheres constantly illuminated by the sun, with a highly oblique spin axis pointing to the sun (aspect angle close to 0$^{\circ}$) and with a near-pole jet (not visible in our images) observed with a small phase angle from Earth. Under this scenario any change in the activity of the comet happened in the same part of the surface, and was not due to diurnal variations (\cite{2013ApJ...779L...3L}). These authors also state that it is unlikely, under this particular geometry, to have diurnal modulation of activity observable as periodic brightness variations. Nevertheless, we obtained periodic brightness variations with a very low peak-to-peak amplitude which can be consistent with this scenario if the pole is not exactly located at the subsolar point. This would produce periodic exposure to sunlight of small parts of the nucleus compatible with the very low amplitude we get from our data. So it is possible to obtain very low photometric variability related to the rotation of the nucleus under this view. 

Finally, a double-peaked light curve is expected for a non-spherical active nucleus if the sublimation comes from the whole surface (i.e. if the object is homogeneously covered of fresh ice). In particular, if variability was mainly due to shape we might expect different heights for the two peaks of the double light curve because a small irregular object does not usually show the same cross section at its maxima. A double-peaked light curve with maxima or minima of different magnitude is not observed. This result also supports the active region-driven variability (single-peaked light curve) against the shape-driven variability (double-peaked light curve). Last, but not least, from photometric analysis of the inner coma of comet C/1995 O1 Hale--Bopp, \cite{1997EM&P...77..207O} obtained a clear synodic single-peaked rotational period of the nucleus of that comet. By similarity with this result --we used the same photometric technique and periodogram analysis for the same type of object, a new comet from the Oort cloud-- we also consider that our obtained period for ISON can be interpreted in the same way: it is the synodic single-peaked rotational period of the ISON nucleus, P = 14.4 $\pm$ 1.2 h.


\section{Summary}
\label{summary}

Time series photometry of the inner coma of comet C/2012 S1 (ISON) obtained when it was at $r_{h} =$ 4.8 AU, $\Delta=$ 4.0 AU on February 8-9 and 11-14, 2013, are presented and analysed in this work.

We have determined a possible synodic rotational period for the nucleus of comet ISON by means of precise photometry of the inner coma and subsequent analysis searching for periodicities. 

Our preferred period from the Lomb-Scargle periodogram analysis is 14.4 $\pm$ 1.2 hours (1.67 cycles/day). Although two apparent 24 h aliases of this main periodicity (9.0 h / 35.7 h) are possible solutions as well. The double-peaked period (28.8 h) is also possible, but less likely than the single-peaked solution. The amplitude of the variability is $0.03\pm0.02$ mag.

All these results seem roughly compatible with the derived geometrical configuration and illumination of ISON's nucleus during February 2013  (\cite{2013ApJ...779L...3L}; \cite{moreno2014}).


\begin{acknowledgements}

This research was based on data obtained at La Hita observatory which is jointly operated by Astrohita and IAA-CSIC. P. Santos-Sanz would like to acknowledge financial support by the Spanish grants AYA2008-06202-C03-01, AYA2011-30106-C02-01, 2007-FQM2998 and 2012-FQM1776. R. Duffard acknowledges financial support from the MICINN (contract Ramon y Cajal).

\end{acknowledgements}



\onecolumn

\begin{longtable}{cccccc}

\caption{Photometry results. \textbf{JD} are the Julian dates corrected for light time at the middle of the exposure for each image; \textbf{R} are the R magnitudes computed as explained in Sect. \ref{results}. The typical error in these R magnitudes estimations is $\sim$ 0.40 mag, but the relative error in the photometry is $\sim$ 0.02 mag; \textbf{m$_R(1,1)$} are ISON's R magnitudes corrected for heliocentric and geocentric distances (not for phase angle). These are the magnitudes that ISON would have at 1 AU from the Sun and the Earth, respectively; \textbf{$r_h$:} heliocentric distance (AU); \textbf{$\Delta$:} geocentric distance (AU); \textbf{$\alpha$:} phase angle (degrees).} \\

\hline 
\hline

JD 	    &	R       &	m$_R(1,1)$   &	$r_h$   & $\Delta	$	&	$\alpha$  \\
	&	[mag]	&	[mag]	       &	[AU]	&	[AU]	    &   [degrees] \\
\hline
\endfirsthead  

\multicolumn{6}{c}{\tablename\ \thetable\ -- Photometry results. \textit{Continued from previous page}} \\

\hline
\hline
JD 	    &	R       &	m$_R(1,1)$   &	$r_h$   & $\Delta	$	&	$\alpha$  \\
	&	[mag]	&	[mag]	       &	[AU]	&	[AU]	    &   [degrees] \\
\hline
\endhead
\hline \multicolumn{6}{c}{\textit{Continued on next page}} \\
\endfoot
\hline
\endlastfoot

\label{photomresults}

2456332.47479 & 16.254 &  9.814 & 4.840 & 4.010 & 6.950 \\
2456332.47681 & 16.251 &  9.811 & 4.840 & 4.010 & 6.950 \\
2456332.47881 & 16.253 &  9.813 & 4.840 & 4.010 & 6.951 \\
2456332.48083 & 16.269 &  9.829 & 4.840 & 4.010 & 6.951 \\
2456332.48286 & 16.279 &  9.839 & 4.840 & 4.010 & 6.951 \\
2456332.48488 & 16.264 &  9.824 & 4.840 & 4.010 & 6.952 \\
2456332.48689 & 16.280 &  9.840 & 4.840 & 4.010 & 6.952 \\
2456332.48891 & 16.264 &  9.824 & 4.840 & 4.010 & 6.953 \\
2456332.51314 & 16.234 &  9.794 & 4.839 & 4.010 & 6.958 \\
2456332.51716 & 16.236 &  9.796 & 4.839 & 4.010 & 6.959 \\
2456332.51919 & 16.255 &  9.815 & 4.839 & 4.010 & 6.959 \\
2456332.52119 & 16.251 &  9.811 & 4.839 & 4.010 & 6.960 \\
2456332.52323 & 16.279 &  9.839 & 4.839 & 4.010 & 6.960 \\
2456332.52523 & 16.286 &  9.847 & 4.839 & 4.010 & 6.961 \\
2456332.52726 & 16.278 &  9.839 & 4.839 & 4.010 & 6.961 \\
2456332.52928 & 16.231 &  9.792 & 4.839 & 4.010 & 6.961 \\
2456332.53130 & 16.251 &  9.812 & 4.839 & 4.010 & 6.962 \\
2456332.53332 & 16.278 &  9.839 & 4.839 & 4.010 & 6.962 \\
2456332.53534 & 16.250 &  9.811 & 4.839 & 4.010 & 6.963 \\
2456332.53735 & 16.267 &  9.828 & 4.839 & 4.010 & 6.963 \\
2456332.53938 & 16.240 &  9.801 & 4.839 & 4.010 & 6.964 \\
2456332.54139 & 16.256 &  9.817 & 4.839 & 4.010 & 6.964 \\
2456332.54339 & 16.251 &  9.812 & 4.839 & 4.010 & 6.964 \\
2456332.54541 & 16.266 &  9.827 & 4.839 & 4.010 & 6.965 \\
2456332.54742 & 16.278 &  9.839 & 4.839 & 4.010 & 6.965 \\
2456332.54946 & 16.269 &  9.830 & 4.839 & 4.010 & 6.966 \\
2456332.55147 & 16.266 &  9.827 & 4.839 & 4.010 & 6.966 \\
2456332.55348 & 16.255 &  9.816 & 4.839 & 4.010 & 6.967 \\
2456332.55551 & 16.234 &  9.795 & 4.839 & 4.010 & 6.967 \\
2456332.55752 & 16.242 &  9.803 & 4.839 & 4.010 & 6.968 \\
2456332.55955 & 16.244 &  9.805 & 4.839 & 4.010 & 6.968 \\
2456332.56159 & 16.236 &  9.797 & 4.839 & 4.010 & 6.968 \\
2456332.56359 & 16.221 &  9.782 & 4.839 & 4.010 & 6.969 \\
2456332.56559 & 16.233 &  9.794 & 4.839 & 4.010 & 6.969 \\
2456332.56763 & 16.251 &  9.812 & 4.839 & 4.010 & 6.970 \\
2456332.56965 & 16.253 &  9.814 & 4.839 & 4.010 & 6.970 \\
2456332.57169 & 16.248 &  9.809 & 4.839 & 4.010 & 6.971 \\
2456332.57372 & 16.261 &  9.822 & 4.839 & 4.010 & 6.971 \\
2456332.57573 & 16.250 &  9.811 & 4.839 & 4.010 & 6.971 \\
2456332.57774 & 16.271 &  9.832 & 4.839 & 4.010 & 6.972 \\
2456332.57975 & 16.286 &  9.847 & 4.839 & 4.010 & 6.972 \\
2456332.58176 & 16.228 &  9.789 & 4.839 & 4.010 & 6.973 \\
2456333.37477 & 16.227 &  9.792 & 4.830 & 4.009 & 7.140 \\
2456333.37679 & 16.232 &  9.797 & 4.830 & 4.009 & 7.141 \\
2456333.37882 & 16.213 &  9.778 & 4.830 & 4.009 & 7.141 \\
2456333.38084 & 16.217 &  9.782 & 4.830 & 4.009 & 7.142 \\
2456333.38287 & 16.231 &  9.796 & 4.830 & 4.009 & 7.142 \\
2456333.38490 & 16.245 &  9.810 & 4.830 & 4.009 & 7.142 \\
2456333.38691 & 16.242 &  9.807 & 4.830 & 4.009 & 7.143 \\
2456333.38895 & 16.220 &  9.785 & 4.830 & 4.009 & 7.143 \\
2456333.39096 & 16.234 &  9.799 & 4.830 & 4.009 & 7.144 \\
2456333.39299 & 16.243 &  9.808 & 4.830 & 4.009 & 7.144 \\
2456333.39501 & 16.236 &  9.801 & 4.830 & 4.009 & 7.145 \\
2456333.39703 & 16.212 &  9.777 & 4.830 & 4.009 & 7.145 \\
2456333.42160 & 16.226 &  9.791 & 4.829 & 4.009 & 7.150 \\
2456333.42359 & 16.230 &  9.795 & 4.829 & 4.009 & 7.151 \\
2456333.42560 & 16.256 &  9.821 & 4.829 & 4.009 & 7.151 \\
2456333.42763 & 16.240 &  9.805 & 4.829 & 4.009 & 7.151 \\
2456333.42963 & 16.237 &  9.803 & 4.829 & 4.009 & 7.152 \\
2456333.43166 & 16.233 &  9.799 & 4.829 & 4.009 & 7.152 \\
2456333.43366 & 16.235 &  9.801 & 4.829 & 4.009 & 7.153 \\
2456333.43568 & 16.246 &  9.812 & 4.829 & 4.009 & 7.153 \\
2456333.43770 & 16.252 &  9.818 & 4.829 & 4.009 & 7.154 \\
2456333.43970 & 16.254 &  9.820 & 4.829 & 4.009 & 7.154 \\
2456333.44171 & 16.253 &  9.819 & 4.829 & 4.009 & 7.155 \\
2456333.44374 & 16.250 &  9.816 & 4.829 & 4.009 & 7.155 \\
2456333.44574 & 16.236 &  9.802 & 4.829 & 4.009 & 7.155 \\
2456333.44775 & 16.241 &  9.807 & 4.829 & 4.009 & 7.156 \\
2456333.44976 & 16.220 &  9.786 & 4.829 & 4.009 & 7.156 \\
2456333.45176 & 16.236 &  9.802 & 4.829 & 4.009 & 7.157 \\
2456333.45376 & 16.218 &  9.784 & 4.829 & 4.009 & 7.157 \\
2456333.45580 & 16.225 &  9.791 & 4.829 & 4.009 & 7.158 \\
2456333.48197 & 16.226 &  9.792 & 4.829 & 4.009 & 7.163 \\
2456333.48399 & 16.247 &  9.813 & 4.829 & 4.009 & 7.164 \\
2456333.48601 & 16.227 &  9.793 & 4.829 & 4.009 & 7.164 \\
2456333.48801 & 16.238 &  9.804 & 4.829 & 4.009 & 7.164 \\
2456333.49001 & 16.240 &  9.806 & 4.829 & 4.009 & 7.165 \\
2456333.49203 & 16.251 &  9.817 & 4.828 & 4.009 & 7.165 \\
2456333.49405 & 16.259 &  9.825 & 4.828 & 4.009 & 7.166 \\
2456333.49606 & 16.251 &  9.817 & 4.828 & 4.009 & 7.166 \\
2456333.49807 & 16.245 &  9.811 & 4.828 & 4.009 & 7.167 \\
2456333.50010 & 16.245 &  9.811 & 4.828 & 4.009 & 7.167 \\
2456333.50212 & 16.226 &  9.792 & 4.828 & 4.009 & 7.167 \\
2456333.50412 & 16.225 &  9.791 & 4.828 & 4.009 & 7.168 \\
2456333.50615 & 16.250 &  9.816 & 4.828 & 4.009 & 7.168 \\
2456333.50816 & 16.241 &  9.807 & 4.828 & 4.009 & 7.169 \\
2456333.51019 & 16.227 &  9.793 & 4.828 & 4.009 & 7.169 \\
2456333.51420 & 16.224 &  9.790 & 4.828 & 4.009 & 7.170 \\
2456333.51623 & 16.260 &  9.826 & 4.828 & 4.009 & 7.170 \\
2456333.51825 & 16.225 &  9.791 & 4.828 & 4.009 & 7.171 \\
2456333.52025 & 16.246 &  9.812 & 4.828 & 4.009 & 7.171 \\
2456333.52226 & 16.270 &  9.836 & 4.828 & 4.009 & 7.172 \\
2456333.52428 & 16.274 &  9.840 & 4.828 & 4.009 & 7.172 \\
2456333.52630 & 16.210 &  9.776 & 4.828 & 4.009 & 7.173 \\
2456333.52832 & 16.244 &  9.810 & 4.828 & 4.009 & 7.173 \\
2456333.53032 & 16.249 &  9.815 & 4.828 & 4.009 & 7.173 \\
2456333.53234 & 16.210 &  9.776 & 4.828 & 4.009 & 7.174 \\
2456333.53434 & 16.249 &  9.815 & 4.828 & 4.009 & 7.174 \\
2456333.53633 & 16.207 &  9.773 & 4.828 & 4.009 & 7.175 \\
2456333.53834 & 16.227 &  9.793 & 4.828 & 4.009 & 7.175 \\
2456333.54037 & 16.248 &  9.814 & 4.828 & 4.009 & 7.176 \\
2456333.54439 & 16.243 &  9.809 & 4.828 & 4.009 & 7.176 \\
2456333.54639 & 16.270 &  9.836 & 4.828 & 4.009 & 7.177 \\
2456333.54841 & 16.260 &  9.826 & 4.828 & 4.009 & 7.177 \\
2456333.55042 & 16.250 &  9.816 & 4.828 & 4.009 & 7.178 \\
2456333.55244 & 16.235 &  9.801 & 4.828 & 4.009 & 7.178 \\
2456333.55446 & 16.213 &  9.779 & 4.828 & 4.009 & 7.179 \\
2456333.55647 & 16.242 &  9.808 & 4.828 & 4.009 & 7.179 \\
2456333.55847 & 16.238 &  9.804 & 4.828 & 4.009 & 7.179 \\
2456333.56047 & 16.230 &  9.796 & 4.828 & 4.009 & 7.180 \\
2456333.56248 & 16.232 &  9.798 & 4.828 & 4.009 & 7.180 \\
2456333.56653 & 16.237 &  9.803 & 4.828 & 4.009 & 7.181 \\
2456333.56855 & 16.238 &  9.804 & 4.828 & 4.009 & 7.182 \\
2456333.57056 & 16.240 &  9.806 & 4.828 & 4.009 & 7.182 \\
2456333.57256 & 16.255 &  9.821 & 4.828 & 4.009 & 7.182 \\
2456333.57457 & 16.262 &  9.828 & 4.828 & 4.009 & 7.183 \\
2456333.57657 & 16.246 &  9.812 & 4.828 & 4.009 & 7.183 \\
2456333.57856 & 16.250 &  9.816 & 4.828 & 4.009 & 7.184 \\
2456333.58059 & 16.234 &  9.800 & 4.828 & 4.009 & 7.184 \\
2456333.58260 & 16.256 &  9.822 & 4.827 & 4.009 & 7.185 \\
2456333.58461 & 16.273 &  9.839 & 4.827 & 4.009 & 7.185 \\
2456333.58662 & 16.278 &  9.844 & 4.827 & 4.009 & 7.185 \\
2456333.58865 & 16.259 &  9.825 & 4.827 & 4.009 & 7.186 \\
2456333.59065 & 16.267 &  9.833 & 4.827 & 4.009 & 7.186 \\
2456333.59265 & 16.265 &  9.831 & 4.827 & 4.009 & 7.187 \\
2456333.59466 & 16.275 &  9.841 & 4.827 & 4.009 & 7.187 \\
2456333.59667 & 16.273 &  9.839 & 4.827 & 4.009 & 7.188 \\
2456333.59869 & 16.239 &  9.805 & 4.827 & 4.009 & 7.188 \\
2456333.60273 & 16.216 &  9.782 & 4.827 & 4.009 & 7.189 \\
2456333.60675 & 16.241 &  9.807 & 4.827 & 4.009 & 7.190 \\
2456333.60877 & 16.263 &  9.829 & 4.827 & 4.009 & 7.190 \\
2456333.61080 & 16.230 &  9.796 & 4.827 & 4.009 & 7.191 \\
2456333.61282 & 16.243 &  9.809 & 4.827 & 4.009 & 7.191 \\
2456333.61485 & 16.215 &  9.782 & 4.827 & 4.009 & 7.191 \\
2456333.61685 & 16.213 &  9.780 & 4.827 & 4.009 & 7.192 \\
2456333.61885 & 16.228 &  9.795 & 4.827 & 4.009 & 7.192 \\
2456333.62087 & 16.232 &  9.799 & 4.827 & 4.009 & 7.193 \\
2456333.62289 & 16.244 &  9.811 & 4.827 & 4.009 & 7.193 \\
2456333.62491 & 16.235 &  9.802 & 4.827 & 4.009 & 7.193 \\
2456335.38203 & 16.272 &  9.848 & 4.808 & 4.008 & 7.560 \\
2456335.38404 & 16.242 &  9.818 & 4.808 & 4.008 & 7.561 \\
2456335.38606 & 16.245 &  9.821 & 4.808 & 4.008 & 7.561 \\
2456335.38808 & 16.245 &  9.821 & 4.807 & 4.008 & 7.561 \\
2456335.39012 & 16.241 &  9.817 & 4.807 & 4.008 & 7.562 \\
2456335.39213 & 16.253 &  9.829 & 4.807 & 4.008 & 7.562 \\
2456335.39417 & 16.241 &  9.817 & 4.807 & 4.008 & 7.563 \\
2456335.39617 & 16.259 &  9.835 & 4.807 & 4.008 & 7.563 \\
2456335.39821 & 16.243 &  9.819 & 4.807 & 4.008 & 7.563 \\
2456335.40022 & 16.249 &  9.825 & 4.807 & 4.008 & 7.564 \\
2456335.40223 & 16.259 &  9.835 & 4.807 & 4.008 & 7.564 \\
2456335.40426 & 16.240 &  9.816 & 4.807 & 4.008 & 7.565 \\
2456335.40626 & 16.252 &  9.828 & 4.807 & 4.008 & 7.565 \\
2456335.44701 & 16.247 &  9.823 & 4.807 & 4.008 & 7.574 \\
2456335.44904 & 16.248 &  9.824 & 4.807 & 4.008 & 7.574 \\
2456335.45106 & 16.250 &  9.826 & 4.807 & 4.008 & 7.575 \\
2456335.48341 & 16.216 &  9.792 & 4.806 & 4.008 & 7.581 \\
2456335.48545 & 16.244 &  9.820 & 4.806 & 4.008 & 7.582 \\
2456335.48748 & 16.239 &  9.815 & 4.806 & 4.008 & 7.582 \\
2456335.48950 & 16.268 &  9.844 & 4.806 & 4.008 & 7.583 \\
2456335.49153 & 16.249 &  9.825 & 4.806 & 4.008 & 7.583 \\
2456335.49354 & 16.270 &  9.846 & 4.806 & 4.008 & 7.584 \\
2456335.49557 & 16.264 &  9.841 & 4.806 & 4.008 & 7.584 \\
2456335.49759 & 16.264 &  9.841 & 4.806 & 4.008 & 7.584 \\
2456335.50164 & 16.254 &  9.831 & 4.806 & 4.008 & 7.585 \\
2456335.50366 & 16.256 &  9.833 & 4.806 & 4.008 & 7.586 \\
2456335.50567 & 16.265 &  9.842 & 4.806 & 4.008 & 7.586 \\
2456335.51376 & 16.246 &  9.823 & 4.806 & 4.008 & 7.588 \\
2456335.51578 & 16.242 &  9.819 & 4.806 & 4.008 & 7.588 \\
2456335.51784 & 16.264 &  9.841 & 4.806 & 4.008 & 7.589 \\
2456335.51986 & 16.268 &  9.845 & 4.806 & 4.008 & 7.589 \\
2456335.52188 & 16.240 &  9.817 & 4.806 & 4.008 & 7.589 \\
2456335.52390 & 16.265 &  9.842 & 4.806 & 4.008 & 7.590 \\
2456335.52596 & 16.255 &  9.832 & 4.806 & 4.008 & 7.590 \\
2456335.52800 & 16.236 &  9.813 & 4.806 & 4.008 & 7.591 \\
2456335.53000 & 16.238 &  9.815 & 4.806 & 4.008 & 7.591 \\
2456335.53203 & 16.272 &  9.849 & 4.806 & 4.008 & 7.592 \\
2456335.53403 & 16.248 &  9.825 & 4.806 & 4.008 & 7.592 \\
2456335.53606 & 16.258 &  9.835 & 4.806 & 4.008 & 7.592 \\
2456335.53811 & 16.241 &  9.818 & 4.806 & 4.008 & 7.593 \\
2456335.54013 & 16.225 &  9.802 & 4.806 & 4.008 & 7.593 \\
2456335.54216 & 16.227 &  9.804 & 4.806 & 4.008 & 7.594 \\
2456335.54419 & 16.198 &  9.775 & 4.806 & 4.008 & 7.594 \\
2456335.54622 & 16.190 &  9.767 & 4.806 & 4.008 & 7.595 \\
2456335.54825 & 16.209 &  9.786 & 4.806 & 4.008 & 7.595 \\
2456335.55027 & 16.209 &  9.786 & 4.806 & 4.008 & 7.595 \\
2456335.55228 & 16.237 &  9.814 & 4.806 & 4.008 & 7.596 \\
2456335.55435 & 16.230 &  9.807 & 4.806 & 4.008 & 7.596 \\
2456335.55635 & 16.243 &  9.820 & 4.806 & 4.008 & 7.597 \\
2456335.55836 & 16.250 &  9.827 & 4.806 & 4.008 & 7.597 \\
2456335.56038 & 16.235 &  9.812 & 4.806 & 4.008 & 7.597 \\
2456335.56240 & 16.232 &  9.809 & 4.806 & 4.008 & 7.598 \\
2456335.56444 & 16.246 &  9.823 & 4.806 & 4.008 & 7.598 \\
2456335.56649 & 16.215 &  9.792 & 4.806 & 4.008 & 7.599 \\
2456335.56851 & 16.233 &  9.810 & 4.805 & 4.008 & 7.599 \\
2456335.57052 & 16.236 &  9.813 & 4.805 & 4.008 & 7.599 \\
2456335.57255 & 16.251 &  9.828 & 4.805 & 4.008 & 7.600 \\
2456335.57459 & 16.243 &  9.820 & 4.805 & 4.008 & 7.600 \\
2456335.57662 & 16.253 &  9.830 & 4.805 & 4.008 & 7.601 \\
2456335.57865 & 16.193 &  9.770 & 4.805 & 4.008 & 7.601 \\
2456335.58067 & 16.228 &  9.805 & 4.805 & 4.008 & 7.602 \\
2456335.58267 & 16.230 &  9.807 & 4.805 & 4.008 & 7.602 \\
2456335.58472 & 16.183 &  9.760 & 4.805 & 4.008 & 7.602 \\
2456335.59685 & 16.185 &  9.762 & 4.805 & 4.008 & 7.605 \\
2456335.60699 & 16.192 &  9.769 & 4.805 & 4.008 & 7.607 \\
2456335.61302 & 16.181 &  9.758 & 4.805 & 4.008 & 7.608 \\
2456335.61506 & 16.219 &  9.796 & 4.805 & 4.008 & 7.609 \\
2456335.61912 & 16.195 &  9.772 & 4.805 & 4.008 & 7.610 \\
2456336.40498 & 16.233 &  9.814 & 4.796 & 4.008 & 7.771 \\
2456336.41103 & 16.193 &  9.774 & 4.796 & 4.008 & 7.772 \\
2456336.41303 & 16.194 &  9.775 & 4.796 & 4.008 & 7.772 \\
2456336.41503 & 16.207 &  9.788 & 4.796 & 4.008 & 7.773 \\
2456336.41706 & 16.203 &  9.784 & 4.796 & 4.008 & 7.773 \\
2456336.41907 & 16.217 &  9.798 & 4.796 & 4.008 & 7.774 \\
2456336.42108 & 16.213 &  9.794 & 4.796 & 4.008 & 7.774 \\
2456336.42310 & 16.204 &  9.785 & 4.796 & 4.008 & 7.775 \\
2456336.42513 & 16.202 &  9.783 & 4.796 & 4.008 & 7.775 \\
2456336.42714 & 16.230 &  9.811 & 4.796 & 4.008 & 7.775 \\
2456336.42917 & 16.210 &  9.791 & 4.796 & 4.008 & 7.776 \\
2456336.43118 & 16.200 &  9.781 & 4.796 & 4.008 & 7.776 \\
2456336.43321 & 16.221 &  9.802 & 4.796 & 4.008 & 7.777 \\
2456336.43521 & 16.198 &  9.779 & 4.796 & 4.008 & 7.777 \\
2456336.43721 & 16.199 &  9.780 & 4.796 & 4.008 & 7.777 \\
2456336.43921 & 16.224 &  9.805 & 4.796 & 4.008 & 7.778 \\
2456336.44124 & 16.195 &  9.776 & 4.796 & 4.008 & 7.778 \\
2456336.44328 & 16.203 &  9.784 & 4.796 & 4.008 & 7.779 \\
2456336.44527 & 16.213 &  9.794 & 4.796 & 4.008 & 7.779 \\
2456336.44729 & 16.199 &  9.780 & 4.796 & 4.008 & 7.780 \\
2456336.44929 & 16.226 &  9.807 & 4.796 & 4.008 & 7.780 \\
2456336.45130 & 16.213 &  9.794 & 4.796 & 4.008 & 7.780 \\
2456336.45333 & 16.209 &  9.790 & 4.796 & 4.008 & 7.781 \\
2456336.45536 & 16.200 &  9.781 & 4.796 & 4.008 & 7.781 \\
2456336.45737 & 16.224 &  9.805 & 4.796 & 4.008 & 7.782 \\
2456336.45940 & 16.208 &  9.789 & 4.796 & 4.008 & 7.782 \\
2456336.46140 & 16.194 &  9.775 & 4.796 & 4.008 & 7.782 \\
2456336.46341 & 16.181 &  9.762 & 4.796 & 4.008 & 7.783 \\
2456336.46543 & 16.178 &  9.759 & 4.796 & 4.008 & 7.783 \\
2456336.46745 & 16.201 &  9.782 & 4.796 & 4.008 & 7.784 \\
2456336.46946 & 16.205 &  9.786 & 4.795 & 4.008 & 7.784 \\
2456336.47147 & 16.212 &  9.793 & 4.795 & 4.008 & 7.785 \\
2456336.47350 & 16.225 &  9.807 & 4.795 & 4.008 & 7.785 \\
2456336.47550 & 16.213 &  9.795 & 4.795 & 4.008 & 7.785 \\
2456336.47750 & 16.217 &  9.799 & 4.795 & 4.008 & 7.786 \\
2456336.47953 & 16.206 &  9.788 & 4.795 & 4.008 & 7.786 \\
2456336.48153 & 16.196 &  9.778 & 4.795 & 4.008 & 7.787 \\
2456336.50172 & 16.197 &  9.779 & 4.795 & 4.008 & 7.791 \\
2456336.50374 & 16.177 &  9.759 & 4.795 & 4.008 & 7.791 \\
2456336.50574 & 16.182 &  9.764 & 4.795 & 4.008 & 7.792 \\
2456337.42286 & 16.243 &  9.829 & 4.785 & 4.008 & 7.978 \\
2456337.42487 & 16.206 &  9.792 & 4.785 & 4.008 & 7.978 \\
2456337.42690 & 16.215 &  9.801 & 4.785 & 4.008 & 7.979 \\
2456337.42890 & 16.212 &  9.798 & 4.785 & 4.008 & 7.979 \\
2456337.43093 & 16.207 &  9.793 & 4.785 & 4.008 & 7.980 \\
2456337.43295 & 16.197 &  9.783 & 4.785 & 4.008 & 7.980 \\
2456337.43495 & 16.192 &  9.778 & 4.785 & 4.008 & 7.980 \\
2456337.43697 & 16.209 &  9.795 & 4.785 & 4.008 & 7.981 \\
2456337.43897 & 16.228 &  9.814 & 4.785 & 4.008 & 7.981 \\
2456337.44100 & 16.211 &  9.797 & 4.785 & 4.008 & 7.982 \\
2456337.44301 & 16.193 &  9.779 & 4.785 & 4.008 & 7.982 \\
2456337.44505 & 16.205 &  9.791 & 4.785 & 4.008 & 7.983 \\
2456337.44706 & 16.213 &  9.799 & 4.785 & 4.008 & 7.983 \\
2456337.44910 & 16.185 &  9.771 & 4.785 & 4.008 & 7.983 \\
2456337.45111 & 16.245 &  9.831 & 4.785 & 4.008 & 7.984 \\
2456337.45314 & 16.227 &  9.813 & 4.785 & 4.008 & 7.984 \\
2456337.45515 & 16.184 &  9.770 & 4.785 & 4.008 & 7.985 \\
2456337.45718 & 16.194 &  9.780 & 4.785 & 4.008 & 7.985 \\
2456337.45918 & 16.200 &  9.786 & 4.784 & 4.008 & 7.985 \\
2456337.46120 & 16.217 &  9.803 & 4.784 & 4.008 & 7.986 \\
2456337.46323 & 16.226 &  9.812 & 4.784 & 4.008 & 7.986 \\
2456337.46525 & 16.204 &  9.790 & 4.784 & 4.008 & 7.987 \\
2456337.46726 & 16.200 &  9.786 & 4.784 & 4.008 & 7.987 \\
2456337.47130 & 16.213 &  9.799 & 4.784 & 4.008 & 7.988 \\
2456337.47332 & 16.174 &  9.761 & 4.784 & 4.008 & 7.988 \\
2456337.47535 & 16.180 &  9.767 & 4.784 & 4.008 & 7.989 \\
2456337.47938 & 16.179 &  9.766 & 4.784 & 4.008 & 7.990 \\
2456337.48138 & 16.219 &  9.806 & 4.784 & 4.008 & 7.990 \\
2456337.48340 & 16.207 &  9.794 & 4.784 & 4.008 & 7.990 \\
2456337.48542 & 16.197 &  9.784 & 4.784 & 4.008 & 7.991 \\
2456337.48745 & 16.204 &  9.791 & 4.784 & 4.008 & 7.991 \\
2456337.48948 & 16.191 &  9.778 & 4.784 & 4.008 & 7.992 \\
2456337.49149 & 16.202 &  9.789 & 4.784 & 4.008 & 7.992 \\
2456337.49351 & 16.220 &  9.807 & 4.784 & 4.008 & 7.992 \\
2456337.49552 & 16.224 &  9.811 & 4.784 & 4.008 & 7.993 \\
2456337.50159 & 16.189 &  9.776 & 4.784 & 4.008 & 7.994 \\
2456337.51169 & 16.179 &  9.766 & 4.784 & 4.008 & 7.996 \\
2456337.51370 & 16.240 &  9.827 & 4.784 & 4.008 & 7.997 \\
2456337.51572 & 16.219 &  9.806 & 4.784 & 4.008 & 7.997 \\
2456337.51774 & 16.202 &  9.789 & 4.784 & 4.008 & 7.997 \\
2456337.51975 & 16.176 &  9.763 & 4.784 & 4.008 & 7.998 \\
2456337.52381 & 16.174 &  9.761 & 4.784 & 4.008 & 7.999 \\
2456337.52583 & 16.201 &  9.788 & 4.784 & 4.008 & 7.999 \\
2456337.52784 & 16.188 &  9.775 & 4.784 & 4.008 & 7.999 \\
2456337.52987 & 16.184 &  9.771 & 4.784 & 4.008 & 8.000 \\
2456337.53190 & 16.175 &  9.762 & 4.784 & 4.008 & 8.000 \\
2456337.53590 & 16.195 &  9.782 & 4.784 & 4.008 & 8.001 \\
2456337.53795 & 16.213 &  9.800 & 4.784 & 4.008 & 8.001 \\
2456337.54199 & 16.214 &  9.801 & 4.784 & 4.008 & 8.002 \\
2456337.54403 & 16.214 &  9.801 & 4.784 & 4.008 & 8.003 \\
2456337.54604 & 16.199 &  9.786 & 4.784 & 4.008 & 8.003 \\
2456337.54804 & 16.194 &  9.781 & 4.784 & 4.008 & 8.004 \\
2456337.55005 & 16.202 &  9.789 & 4.783 & 4.008 & 8.004 \\
2456337.55208 & 16.178 &  9.765 & 4.783 & 4.008 & 8.004 \\
2456337.55410 & 16.177 &  9.764 & 4.783 & 4.008 & 8.005 \\
2456337.55613 & 16.192 &  9.779 & 4.783 & 4.008 & 8.005 \\
2456337.55817 & 16.238 &  9.825 & 4.783 & 4.008 & 8.006 \\
2456337.56021 & 16.221 &  9.808 & 4.783 & 4.008 & 8.006 \\
2456337.56226 & 16.206 &  9.793 & 4.783 & 4.008 & 8.006 \\
2456337.56428 & 16.225 &  9.812 & 4.783 & 4.008 & 8.007 \\
2456337.56628 & 16.210 &  9.797 & 4.783 & 4.008 & 8.007 \\
2456337.56831 & 16.222 &  9.809 & 4.783 & 4.008 & 8.008 \\
2456337.57034 & 16.239 &  9.826 & 4.783 & 4.008 & 8.008 \\
2456337.57237 & 16.185 &  9.772 & 4.783 & 4.008 & 8.009 \\
2456337.57439 & 16.191 &  9.778 & 4.783 & 4.008 & 8.009 \\
2456337.57641 & 16.185 &  9.772 & 4.783 & 4.008 & 8.009 \\
2456337.57845 & 16.204 &  9.791 & 4.783 & 4.008 & 8.010 \\
2456337.58248 & 16.233 &  9.820 & 4.783 & 4.008 & 8.011 \\
2456337.58449 & 16.236 &  9.823 & 4.783 & 4.008 & 8.011 \\
2456337.59057 & 16.180 &  9.767 & 4.783 & 4.008 & 8.012 \\
2456337.59260 & 16.190 &  9.777 & 4.783 & 4.008 & 8.013 \\
2456337.59463 & 16.191 &  9.778 & 4.783 & 4.008 & 8.013 \\
2456337.59667 & 16.201 &  9.788 & 4.783 & 4.008 & 8.013 \\
2456337.59870 & 16.205 &  9.792 & 4.783 & 4.008 & 8.014 \\
2456337.60072 & 16.202 &  9.789 & 4.783 & 4.008 & 8.014 \\
2456337.60275 & 16.213 &  9.800 & 4.783 & 4.008 & 8.015 \\
2456338.32476 & 16.224 &  9.815 & 4.775 & 4.008 & 8.159 \\
2456338.32677 & 16.229 &  9.820 & 4.775 & 4.008 & 8.159 \\
2456338.32878 & 16.211 &  9.802 & 4.775 & 4.008 & 8.160 \\
2456338.33079 & 16.234 &  9.825 & 4.775 & 4.008 & 8.160 \\
2456338.33282 & 16.236 &  9.827 & 4.775 & 4.008 & 8.161 \\
2456338.33483 & 16.241 &  9.832 & 4.775 & 4.008 & 8.161 \\
2456338.33686 & 16.217 &  9.808 & 4.775 & 4.008 & 8.162 \\
2456338.36032 & 16.225 &  9.816 & 4.774 & 4.008 & 8.166 \\
2456338.36238 & 16.234 &  9.825 & 4.774 & 4.008 & 8.167 \\
2456338.36441 & 16.208 &  9.799 & 4.774 & 4.008 & 8.167 \\
2456338.36645 & 16.212 &  9.803 & 4.774 & 4.008 & 8.168 \\
2456338.36845 & 16.190 &  9.781 & 4.774 & 4.008 & 8.168 \\
2456338.37047 & 16.206 &  9.797 & 4.774 & 4.008 & 8.168 \\
2456338.37250 & 16.230 &  9.821 & 4.774 & 4.008 & 8.169 \\
2456338.37454 & 16.248 &  9.839 & 4.774 & 4.008 & 8.169 \\
2456338.37655 & 16.217 &  9.808 & 4.774 & 4.008 & 8.170 \\
2456338.37859 & 16.226 &  9.817 & 4.774 & 4.008 & 8.170 \\
2456338.38060 & 16.217 &  9.808 & 4.774 & 4.008 & 8.170 \\
2456338.38262 & 16.247 &  9.838 & 4.774 & 4.008 & 8.171 \\
2456338.38463 & 16.243 &  9.834 & 4.774 & 4.008 & 8.171 \\
2456338.38667 & 16.195 &  9.786 & 4.774 & 4.008 & 8.172 \\
2456338.38868 & 16.202 &  9.793 & 4.774 & 4.008 & 8.172 \\
2456338.39072 & 16.196 &  9.787 & 4.774 & 4.008 & 8.172 \\
2456338.39272 & 16.232 &  9.823 & 4.774 & 4.008 & 8.173 \\
2456338.39476 & 16.224 &  9.815 & 4.774 & 4.008 & 8.173 \\
2456338.39678 & 16.207 &  9.798 & 4.774 & 4.008 & 8.174 \\
2456338.39880 & 16.200 &  9.791 & 4.774 & 4.008 & 8.174 \\
2456338.40083 & 16.241 &  9.832 & 4.774 & 4.008 & 8.174 \\
2456338.40286 & 16.226 &  9.817 & 4.774 & 4.008 & 8.175 \\
2456338.40490 & 16.234 &  9.825 & 4.774 & 4.008 & 8.175 \\
2456338.40693 & 16.227 &  9.818 & 4.774 & 4.008 & 8.176 \\
2456338.40896 & 16.224 &  9.815 & 4.774 & 4.008 & 8.176 \\
2456338.41097 & 16.208 &  9.799 & 4.774 & 4.008 & 8.177 \\
2456338.41297 & 16.211 &  9.802 & 4.774 & 4.008 & 8.177 \\
2456338.41502 & 16.227 &  9.818 & 4.774 & 4.008 & 8.177 \\
2456338.41704 & 16.234 &  9.825 & 4.774 & 4.008 & 8.178 \\
2456338.41904 & 16.223 &  9.814 & 4.774 & 4.008 & 8.178 \\
2456338.42106 & 16.203 &  9.794 & 4.774 & 4.008 & 8.179 \\
2456338.42308 & 16.177 &  9.768 & 4.774 & 4.008 & 8.179 \\
2456338.49481 & 16.221 &  9.812 & 4.773 & 4.008 & 8.193 \\
2456338.49684 & 16.217 &  9.808 & 4.773 & 4.008 & 8.194 \\
2456338.49888 & 16.236 &  9.827 & 4.773 & 4.008 & 8.194 \\
2456338.50089 & 16.225 &  9.816 & 4.773 & 4.008 & 8.195 \\
2456338.50293 & 16.221 &  9.813 & 4.773 & 4.008 & 8.195 \\
2456338.50497 & 16.243 &  9.835 & 4.773 & 4.008 & 8.196 \\
2456338.50701 & 16.215 &  9.807 & 4.773 & 4.008 & 8.196 \\
2456338.50902 & 16.193 &  9.785 & 4.773 & 4.008 & 8.196 \\
2456338.51103 & 16.214 &  9.806 & 4.773 & 4.008 & 8.197 \\
2456338.51306 & 16.243 &  9.835 & 4.773 & 4.008 & 8.197 \\
2456338.51508 & 16.216 &  9.808 & 4.773 & 4.008 & 8.198 \\
2456338.51708 & 16.186 &  9.778 & 4.773 & 4.008 & 8.198 \\
2456338.51913 & 16.208 &  9.800 & 4.773 & 4.008 & 8.198 \\
2456338.52116 & 16.225 &  9.817 & 4.773 & 4.008 & 8.199 \\
2456338.52319 & 16.220 &  9.812 & 4.773 & 4.008 & 8.199 \\
2456338.52521 & 16.231 &  9.823 & 4.773 & 4.008 & 8.200 \\
2456338.52725 & 16.219 &  9.811 & 4.773 & 4.008 & 8.200 \\
2456338.52925 & 16.249 &  9.841 & 4.773 & 4.008 & 8.200 \\
2456338.55550 & 16.195 &  9.787 & 4.772 & 4.008 & 8.206 \\
2456338.55753 & 16.182 &  9.774 & 4.772 & 4.008 & 8.206 \\
2456338.56160 & 16.206 &  9.798 & 4.772 & 4.008 & 8.207 \\
2456338.56362 & 16.231 &  9.823 & 4.772 & 4.008 & 8.207 \\
2456338.56562 & 16.199 &  9.791 & 4.772 & 4.008 & 8.208 \\
2456338.56766 & 16.210 &  9.802 & 4.772 & 4.008 & 8.208 \\
2456338.56969 & 16.198 &  9.790 & 4.772 & 4.008 & 8.209 \\
2456338.57169 & 16.215 &  9.807 & 4.772 & 4.008 & 8.209 \\
2456338.57372 & 16.207 &  9.799 & 4.772 & 4.008 & 8.209 \\
2456338.57573 & 16.207 &  9.799 & 4.772 & 4.008 & 8.210 \\
2456338.57777 & 16.197 &  9.789 & 4.772 & 4.008 & 8.210 \\
2456338.57983 & 16.227 &  9.819 & 4.772 & 4.008 & 8.211 \\
2456338.58185 & 16.231 &  9.823 & 4.772 & 4.008 & 8.211 \\
2456338.58389 & 16.225 &  9.817 & 4.772 & 4.008 & 8.211 \\
2456338.58593 & 16.203 &  9.795 & 4.772 & 4.008 & 8.212 \\
2456338.58796 & 16.184 &  9.776 & 4.772 & 4.008 & 8.212 \\
2456338.58998 & 16.212 &  9.804 & 4.772 & 4.008 & 8.213 \\
2456338.59200 & 16.248 &  9.840 & 4.772 & 4.008 & 8.213 \\
2456338.59403 & 16.230 &  9.822 & 4.772 & 4.008 & 8.213 \\
2456338.59606 & 16.187 &  9.779 & 4.772 & 4.008 & 8.214 \\
2456338.60015 & 16.181 &  9.773 & 4.772 & 4.008 & 8.215 \\
2456338.60417 & 16.188 &  9.780 & 4.772 & 4.008 & 8.215 \\

\hline

\end{longtable}

\twocolumn

\end{document}